3000 % high-field magnetoresistance in super-lattices of CoFe nanoparticles.

Reasmey P. Tan[a], Julian Carrey[a], Marc Respaud[a], Céline Desvaux[b], Philippe Renaud[b], Bruno Chaudret[c]

[a] LPCNO-IRSAMC-INSA-UPS-CNRS, 135, av. de Rangueil, 31077 Toulouse cedex, France
[b] Freescale Semiconductor, le Mirail B.P. 1029, 31023 Toulouse Cedex, France
[c] Laboratoire de Chimie de Coordination-CNRS, 205 rte de Narbonne, 31077 Toulouse cedex 4, France

**Abstract :**
We report on magnetotransport measurements on millimetre-large super-lattices of CoFe nanoparticles surrounded by an organic layer. Electrical properties are typical of Coulomb blockade in three dimensional arrays of nanoparticles. A large high-field magnetoresistance, reaching up to 3000 %, is measured between 1.8 and 10 K. This exceeds by two orders of magnitude magnetoresistance values generally measured in arrays of 3d metals ferromagnetic nanoparticles. The magnetoresistance amplitude scales with the magnetic field / temperature ratio and displays an unusual exponential dependency with the applied voltage. The magnetoresistance abruptly disappears below 1.8 K. We propose that the magnetoresistance is due to some individual paramagnetic moments localized between the metallic core of the nanoparticles, the origin of which is discussed.

**Corresponding authors:**
Julian Carrey (julian.carrey@insa-toulouse.fr) and Marc Respaud (respaud@insa-toulouse.fr)
Laboratoire de Physique et Chimie des Nano-Objets
135 av. de Rangueil
31077 Toulouse cedex 4
France
tel : +33 561559677
fax : +33 561559697

**Main text :**

Arrays of magnetic nanoparticles surrounded by insulating layers can display both Coulomb blockade and tunnel magnetoresistance [1]. Interest in their study has been driven by theoretical predictions that the interplay between these two properties could lead to effects such as oscillations in the magnetoresistance (MR) amplitude as a function of the applied voltage [2] or enhancement of the tunnel MR in the Coulomb blockade regime [3]. Several types of structures have been studied [4-16], with magnetic nanoparticles composed of cobalt [4-9], iron [10,11], $Sr_2FeMoO_6$ [16], or magnetite [12-15]. These particles were surrounded by insulating barriers composed of organic materials [4-13], $MgF_2$ [5], gas [6,7], $Al_2O_3$ [8-10], iron oxide [11] or grain boundaries [12,14-16]. In some cases, low-field tunnel MR has been reported [4-10], as well as experimental indications of the influence of Coulomb blockade on the MR properties [8-10]. From the low-field tunnel MR, the spin polarisation of the magnetic nanoparticles can be deduced [5-7,10]. Moreover, MR occurring above the saturation field of the magnetic particles has been reported in a few systems [11-17]. When the particles are composed of 3d metals ferromagnets, these effects have been attributed to the presence of isolated atoms inside the tunnel barrier [17]. When the nanoparticles are composed of magnetic oxides, such effects have been explained by the presence of disordered magnetic moments at their surface [18,19].

In arrays of 3d metals ferromagnetic nanoparticles (Co, Fe, Ni) surrounded by non-magnetic tunnel barriers, the MR ratio generally measured is small, limited to a few tens of percent [4-10]. Indeed, the spin polarisation of these materials and their alloys ranges from 30 to 60 %, which does not lead to large values of magnetoresistance [1]. In this article, magnetotransport properties of three-dimensional arrays of Co-Fe nanoparticles surrounded by insulating organic ligands are described. Huge values of MR, reaching up to 3000 % are measured, which exceeds by two orders of magnitude the values generally measured in arrays of 3d metals ferromagnetic nanoparticles. Thereafter we detail the features of this MR and discuss its possible origin.

The samples we studied are composed of ferromagnetic monodisperse Cobalt-Iron nanoparticles synthesized using organometallic chemistry and organized inside millimeter-long super-lattices with a FCC packing. Their synthesis method and their detailed characterization have been reported elsewhere [20]. The particles are 15 nm in diameter, separated by a 2 nm organic barrier composed of a mixture of hexadecylamine and long chain carboxylic acids (see Fig. 1). A coercive field of 0.03 T and a saturation field of 0.8 T are measured at 2 K on a single superlattice when the magnetic field is applied perpendicularly to the needle. Super-lattices were connected using Au wires and silver paint in a glove box under Ar atmosphere. The typical distance between the two contacts is 0.5 mm with around 30000 nanoparticles measured in series/parallel. The time to transfer the connected sample from the glove box to the inside of the cryostat was kept to a minimum (few tens of seconds) to prevent oxidation. Magnetotransport measurements were performed in a Cryogenic cryostat equipped with a superconducting coil (up to 10 T) using a Keithley 6430 sub-femtoamp sourcemeter. In the following experiments, the magnetic field was applied perpendicularly to the current and to the easy axis of the superlattice, with a constant sweep rate in the range of 0.009 T/s. The resistances of the super-lattices were measured in a range between 8 k$\Omega$ and 5 G$\Omega$ at room temperature. The origin of this broad distribution is still unclear but may be related to a dispersion in the contact resistance or to micro-cracks inside the samples. They display three different types of MR, including low-field MR and an original mechanism of MR, which are detailed elsewhere [21]. In this article, we only focus on the high-field MR properties. High-field MR has been observed on all the samples, and its highest amplitude varies from sample to sample between 80 and 3000 %. All the results presented here arise from the super-lattice displaying the highest high-field MR, but similar behaviour with different MR amplitudes was observed on all the samples we measured.

Resistance-temperature ($R(T)$) and current-voltage ($I(V)$) characteristics of the super-lattice present features typical of Coulomb blockade in an array of nanoparticles. Resistance measured at a voltage of 25 V increases exponentially with decreasing temperature (see Fig. 2a). In an array of nanoparticles in the Coulomb blockade regime, the low-bias resistance $R$ follows $R = R_O \exp[(T_O/T)^\nu]$, where $R_0$ is the high-temperature resistance and $T_0$ the activation temperature for charge transport in the array [22]. $\nu = 1/2$ or $\nu = 1$ are generally observed experimentally. Fig. 2 shows that $\ln(R)$ varies linearly with $T^{-1/2}$ between 2 K and 100 K. From the slope of the curve, the activation energy $T_0 = 66\ (\pm 2)$ K is extracted.

$I(V)$ characteristics measured at various temperatures are presented in Fig. 2b. The Coulomb gap progressively opens when the temperature is lowered. In arrays of nanoparticles, $I(V)$ characteristics in the Coulomb blockade regime are expected to follow at low temperature the relation $I \propto [(V - V_T)/V_T]^\zeta$, where $V_T$ is the threshold voltage above which conduction occurs. $\zeta$ is an exponent related to the dimensionality: $\zeta = 1$ and $\zeta = 5/3$ have been calculated for 1D and 2D arrays of disordered nanoparticles respectively [23-25]. Even if no theoretical value is available for 3D arrays, larger exponents are expected and measured [26]. Our lowest temperature $I(V)$ characteristic at 1.7 K is well fitted by the previous equation (see Fig. 2b). $\zeta = 3.75\ (\pm 0.3)$ is determined from the slope of the $I.dV/dI$ vs $V$ curve, and $V_T = 24\ (\pm 5)$ V is then determined by a least squares fitting of the $I(V)$ characteristic with $\zeta$ as a fixed parameter [23]. The value for $\zeta$ is comparable to the one measured by Lebreton *et al.* on thick arrays of Pd clusters [26], which confirms the three-dimensional nature of the electrical conduction in the superlattices.

We now turn to high-field MR properties. In the following, the MR is defined as $(R_h - R_l) / R_l$, where $R_h$ ($R_l$) is the high (low) resistance of the $R(H)$ characteristic. Between 1.8 and 10 K, a high-field MR is observed (see Fig. 3c for a typical $R(H)$ curve). The peaks appearing at $\mu_0 H = 0.5$ T and the butterfly shape of the $R(H)$ characteristic are both due to dynamical effects : when the resistance of the sample is measured as a function of time after varying the magnetic field, a lag of about 90 seconds between the variation of magnetic field and the variation of resistance is observed, but this is in no case related to inductive currents inside the sample. Fig. 3a displays the evolution of the MR at 2.6 T as a function of temperature measured for $V = 200$ V. Two well-defined regimes are observed. In the first regime, between 2.4 and 10 K, the amplitude of the high-field MR regularly increases when the temperature is lowered. The MR amplitude strongly depends on the applied voltage. At 3.15 K, its amplitude at $\mu_0 H = 8.8$ T varies from 41 % at 200 V to 3000 % at 20 V (see Fig 3b and 3c). The complete MR($V$) dependence is extracted from two $I(V)$ characteristics measured at $\mu_0 H = 0$ T and $\mu_0 H = 8.8$ T. The MR($V$) is given by $[I(V, \mu_0 H = 8.8\ T) - I(V, \mu_0 H = 0\ T)] / I(V, \mu_0 H = 0\ T)$. An unusual exponential dependence of the MR is observed as a function of the applied voltage (see Fig. 3d). Interestingly, all MR curves measured between 3.1 and 10 K superpose perfectly when plotted as a function of $H/T$ (see Fig. 4). The MR amplitude displays a plateau below 2.4 K and abruptly collapses below 1.8 K (see Fig. 3a). In this second regime, below 1.8 K, the amplitude of the MR is only 2 % at 200 V and 2.6 T.

This high-field MR cannot be attributed to simple tunnel magnetoresistance between spin-polarized nanoparticles. Indeed, even when assuming a full spin polarisation of the NPs, the maximum tunnel magnetoresistance would be of only 100 %, assuming a random orientation of the anisotropy axis of the nanoparticles [1]. There are a few indications that the MR in this system could be due to paramagnetic species : i) The $H/T$ dependence is typical of a Langevin-like behaviour corresponding to the magnetization of paramagnetic or superparamagnetic species. However, the non-saturation of the MR up to 10 T at 2 K is not compatible with the presence of superparamagnetic impurities which would be saturated in few T. ii) Below 1.8 K, when the high-field MR disappears, the $R(H)$ curve has a typical shape of inverse tunnelling MR characteristic [21]. This type of MR appears when the transport occurs via impurities or states localized within the insulating barrier [27]. iii) Moreover, the abrupt transition between the two regimes below 1.8 K may be related to a paramagnetic-ferromagnetic or a paramagnetic-glass transition of these localized magnetic moments.

We do not have any experimental proof of the presence of these isolated magnetic moments but a few hypothesis can be made to explain their eventual presence. Localized states with a paramagnetic-like magnetic moment can be induced at the surface of the nanoparticles by the surface coordination of carboxylic acids, which are known to destabilize the magnetism of surface atoms by an electronic transfer from the particle to the molecule. It has already been experimentally shown that the bounding of molecules on metal surfaces can induce such states [28]. A localized surface state could also be induced by a small uncontrolled oxidation of the particles surface during the transfer between the glove box and the cryostat in spite of all the precautions taken. Finally, a localized state can be induced by some remaining traces of organometallic complexes or clusters of Co or Fe formed during the synthesis. The experimental discrimination between these hypotheses is not trivial and deserves much more work. Indeed, magnetization measurements up to 5 T in the same range of temperature do not allow us to conclude unambiguously. The differential high field susceptibility is bigger than in bulk materials. Such a larger value could be related either to a disordered surface state or to the presence of some magnetic impurities.

We now discuss the way that these impurities could lead to such large values of MR. Huang *et al.* have shown that the large values of MR observed in nanoparticles with a disordered insulating magnetic surface may be explained by modeling the surface as a supplementary effective tunnel barrier, the height of which is lowered by a magnetic field [18]. We have tried to a apply such a phenomenological model to fit our observations. Preliminary results indicate this could reproduce the large amplitude of MR and its strong voltage dependence. However, a detailed presentation of this model is beyond the scope of this Letter, which is aimed only to present the experimental results.

In conclusion, we have investigated the magnetotransport properties of three-dimensional super-lattices of ferromagnetic FeCo nanoparticles. The electronic transport characteristics are fitted with equations verified in arrays of nanoparticles in the Coulomb blockade regime. Our results show that large values of magnetoresistance can be obtained in arrays of 3d ferromagnetic metals nanoparticles. The MR is proposed to be due to the presence of paramagnetic defects localised either on the surface of the nanoparticle or within the organic barrier.

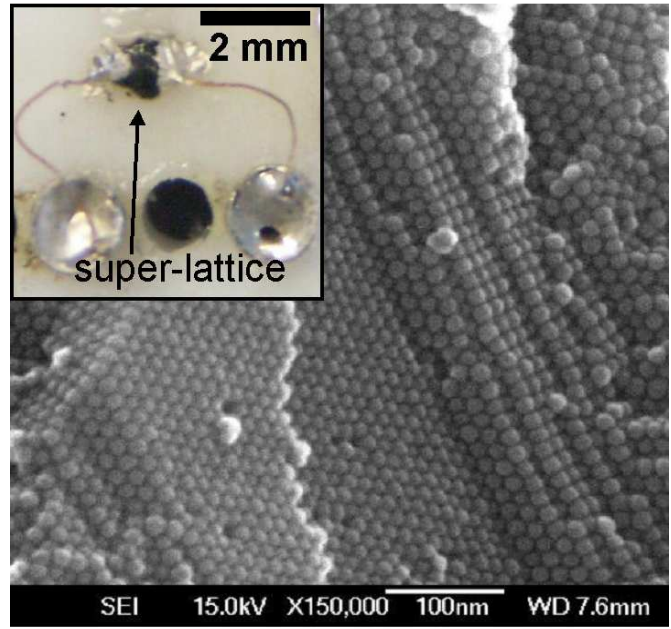

Fig. 1 : Scanning electron microscope micrograph of a typical millimetre-long fcc super-lattice composed of 15 nm spherical Cobalt-Iron nanoparticles. The micrograph shows a broken part of a superlattice so the inner organisation is visible (inset) Picture of a connected superlattice. The sample holder, gold wires, silver painting and the superlattice itself are visible.

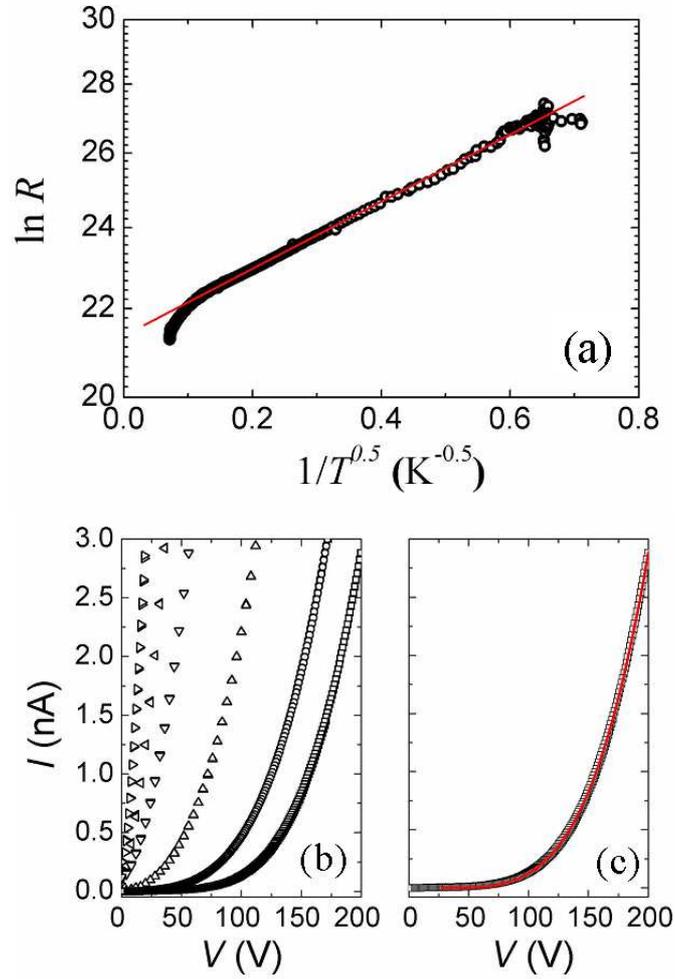

Fig 2: (a) ln (*R*) plotted as a function of $T^{-1/2}$, measured at 25 V. The linear slope between 2 and 100 K is used to deduce $T_0$ = 66 K. (b) *I*(*V*) characteristics measured at (from left to right) 26, 11, 7.5, 3.7, 2.7 and 1.7 K. (c) fit of the *I*(*V*) characteristic at 1.7 K, with $\zeta$ = 3.75 and $V_T$ = 24 V.

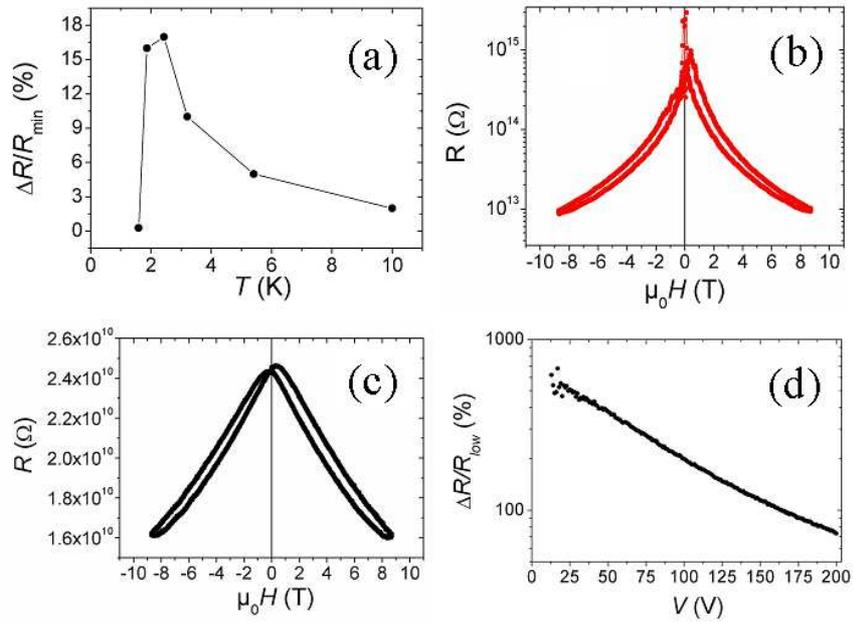

Fig. 3: (a) Magnetoresistance amplitude as a function of temperature, measured at 2.6 T for $V = 200$ V. (b) $R(H)$ characteristic measured at $V = 20$ V and $T = 3.15$ K (logarithmic scale). (c) $R(H)$ characteristic measured at $V = 200$ V and $T = 3.15$ K (linear scale). (d) Voltage dependence of the magnetoresistance at 2.75 K and 8.8 T (logarithmic scale), calculated from two $I(V)$ characteristics measured at 0 and 8.8 T.

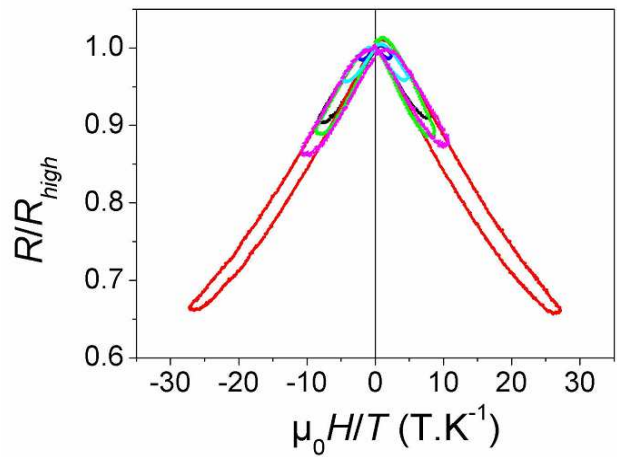

Fig. 4 : Magnetoresistance characteristics plotted as a function of $H/T$. They are measured at various temperatures between 3.1 and 10 K and various maximum magnetic fields between 1.5 and 8.6 T, for $V$ = 200 V.